\documentclass[aps,pre,preprint,groupedaddress,showpacs]{revtex4}
\usepackage{amssymb}
\usepackage{graphicx}

\begin{document}

\title{Slowdown and Splitting of Gap Solitons in Apodized Bragg Gratings}
\author{William C. K. Mak$^{1}$, Boris A. Malomed$^{2,1}$, and Pak L. Chu$^1$%
}
\affiliation{$^1$Optoelectronic Research Centre, Department of Electronic Engineering,
City University of Hong Kong}
\affiliation{$^2$Department of Interdisciplinary Studies, Faculty of Engineering, Tel
Aviv University, Tel Aviv 69978, Israel\thanks{
permanent address}}

\begin{abstract}
We study the motion of gap solitons in two models of \textit{apodized}
nonlinear fiber Bragg gratings (BGs), with the local reflectivity $\kappa$
varying along the fiber. A single step of $\kappa $, and a periodic array of
alternating steps with opposite signs (a ``Bragg superstructure") are
considered. These structures may be used in the design of various optical
elements employing the gap solitons. A challenging possibility is to slow
down and eventually halt the soliton by passing it through the step of
increasing reflectivity, thus capturing a pulse of \emph{standing light}.
First, we develop an analytical approach, assuming adiabatic evolution of
the soliton, and making use of the energy conservation and balance equation
for the momentum. Comparison with simulations shows that the analytical
approximation is quite accurate, unless the inhomogeneity is too narrow, or
the step is too high: the soliton is either transmitted across the step or
bounces back from it. If the step is narrow, systematic simulations
demontrate that the soliton \emph{splits} into transmitted and reflected
pulses (splitting of a BG soliton which hits a chirped grating was observed
in experiments). Moving through the periodic ``superstructure'' , the
soliton accummulates distortion and suffers radiation loss if the structure
is composed of narrow steps. The soliton moves without any loss or
irrieversible deformation through the array of sufficiently broad steps.
\end{abstract}

\maketitle

\section{Introduction}

Bragg gratings (BGs) are periodic structures produced by a periodic
variation of the refractive index along a fiber or an optical waveguide. 
Devices based on fiber gratings are widely used in optical
systems \cite{Kashyap}. \textit{Gap solitons} (in this context, they are
also called BG solitons \cite{Sterke}) are supported by fiber gratings
through the balance between the BG-induced linear dispersion (which includes
a \textit{gap} in the system's linear spectrum) and Kerr nonlinearity of the
fiber material. Analytical solutions for BG solitons in the standard
fiber-grating model are well known \cite{Aceves,Demetri}. Studies of the
stability of these solutions have led to a conclusion that almost exactly
half of them are stable (see details below) \cite{Tasgal}. These solitons
were created and studied in detail in the experiment \cite%
{experiment,slowest}.

Recently, much attention has been attracted to the creation of ``slow light"
in various media \cite{slowlight}, including fibers (for instance, a thin
fiber embedded in an electromagnetically controlled molecular solid \cite%
{slowlightJModPhys}). In particular, the possibilities to generate slowly moving
optical solitons were considered in various settings \cite%
{slowsoliton,we_collision}. The nonlinear fiber grating is a medium where it
may be possible to stop a soliton. Indeed, solutions for \emph{zero-velocity}
gap solitons are available, in which the counter-propagating waves are
mutually locked in the dynamical equilibrium between the linear
Bragg-resonant conversion and Kerr nonlinearity keeping them together
through the XPM (cross-phase-modulation) interaction \cite{Aceves,Demetri}.

Actually, the BG solitons that were created in the experiments to date were
fast ones, the slowest among them having the velocity equal to half of the
light velocity $\overline{c}$ in the fiber \cite{slowest}. A possibility to
capture a standing BG soliton is to bind it on an attractive local
inhomogeneity \cite{Weinstein,we} (earlier, it was demonstrated that a local
defect in the fiber grating can also stimulate a nonlinear four-wave
interaction without formation of a soliton \cite{trapping}). Moreover, it is
possible to combine the attractive inhomogeneity and local linear gain,
which is a way to create a permanently pinned soliton in the presence of
loss \cite{we2}. Stable pinned-solitons are not only objects of fundamental
interest, but they also have an obvious potential for design of various
nonlinear-optical elements.

Another approach to the creation of standing BG solitons was explored in a
recent work \cite{we_collision}, where it was demonstrated that head-on
collisions between two moving solitons may lead to their fusion into an
immobile soliton (with residual internal oscillations, which is explained by
the existence of an oscillatory intrinsic mode in the stable BG soliton \cite%
{Tasgal}), provided that their initial velocities are smaller than $\approx
0.2~\overline{c}$. Thus, one should still bridge the remaining gap between
the experimentally available minimum soliton velocity, $0.5\overline{c}$, and
the necessary value of $0.2\overline{c}$.

A possibility to resolve this problem was also proposed and briefly
considered in Ref. \cite{we_collision} (a similar possibility was mentioned
still earlier in Ref. \cite{express}): passing the BG soliton through an 
\textit{apodized} fiber grating, with the Bragg reflectivity gradually
increasing along the fiber. In this case, the soliton may be adiabatically
slowed down. An objective of the present work is to consider this problem in
a systematic form, both analytically and numerically. Besides that, we also
aim to study the motion of the soliton through a periodic array of
alternating reflectivity steps with opposite signs.

It should be mentioned that apodized fiber gratings (a gradually varying
reflectivity may also be featured by \textit{chirped} gratings \cite%
{express,Tsoy}) are widely used in various applications \cite{Kashyap}. Some
experimental \cite{slowest, express} and theoretical \cite{Tsoy} results
concerning passage of BG solitons through apodized gratings were reported
earlier. These include the use of the apodization to facilitate the launch
of solitons into the fiber grating \cite{slowest}, observation of splitting
of a BG soliton which hits a chirped grating \cite{express}, and analysis of
the soliton's dynamics in the case when the apodized-grating model reduces
to a perturbed nonlinear Schr\"{o}dinger equation \cite{Tsoy}. The
formulation of problems considered in this work, and methods developed for
the analysis, are essentially different from the earlier works, as will be
seen below.

The rest of the paper is organized as follows. The model is formulated in
section 2, and in section 3 we work out the analytical approach, which is
based on treating the gap soliton moving in the apodized BG as a
quasi-particle obeying an equation of motion derived from the balance
equation for the soliton's momentum. This adiabatic approach is relevant
when the Bragg reflectivity varies on a spatial scale which exceeds the size
of the soliton (for slow solitons, a physically relevant size is, typically, 
$\sim 1$ mm, while for fast solitons the size is additionally subject to the
``Lorentz contraction", see below). Results of systematic numerical
simulations of the full model with a step of the Bragg reflectivity (that
will also be sometimes called \textquotedblleft barrier'' ) are presented in
section 4. First, we compare the analytical predictions produced by the
adiabatic approximation with numerical results, and conclude that the
adiabatic approximation is quite accurate, provided that the step is not too
steep. Further, we present numerical results for the case of steeper
apodization. In that case, the soliton impinging upon the step may split
into two secondary solitons (transmitted and reflected ones), or get
completely destroyed. In section 5 we briefly describe results of
simulations of the motion of the soliton through a periodic array of
alternating positive and negative steps. It is found that the soliton
distorts itself and emits radiation if the steps are narrow. If they are
broader, the deformation of the moving soliton accumulates slower, and it
can move without any irreversible damage through the array of sufficiently
broad steps. Section 6 concludes the paper.

\section{The model}

The commonly adopted model of the nonlinear fiber grating is based on a
system of coupled equations for the local amplitudes of the right- ($u$) and
left- ($v$) traveling waves \cite{Sterke},

\begin{eqnarray}
iu_{t}+iu_{x}+\kappa v+\left( |u|^{2}/2+|v|^{2}\right) u &=&0,  \nonumber \\
iv_{t}-iv_{x}+\kappa u+\left( |v|^{2}/2+|u|^{2}\right) v &=&0,  \label{pdes}
\end{eqnarray}
where $x$ and $t$ are the coordinate and time, which are scaled so that the
linear group velocity of light in the fiber is $1$, and $\kappa $ is the
Bragg-reflectivity coefficient. In the apodized grating, $\kappa $ is a
function of $x$.

Exact solutions to Eqs. (\ref{pdes}) with $\kappa \equiv \mathrm{const}$,
which describe solitons moving at a velocity $c$ ($c^{2}<1$) through the
uniform BG, were found in Refs. \cite{Aceves} and \cite{Demetri}: 
\begin{eqnarray}
u_{\mathrm{sol}} &=&\sqrt{\frac{2\kappa \left( 1+c\right) }{3-c^{2}}}\left(
1-c^{2}\right) ^{1/4}W(X)\mathrm{exp}\,\left[ i\phi (X)-iT\cos \theta \right]
,  \nonumber \\
v_{\mathrm{sol}} &=&-\sqrt{\frac{2\kappa \left( 1-c\right) }{3-c^{2}}}\left(
1-c^{2}\right) ^{1/4}W^{\ast }(X)\mathrm{exp}\,\left[ i\phi (X)-iT\cos
\theta \right] .  \label{movsol}
\end{eqnarray}
\noindent Here, the asterisk stands for the complex conjugation, $\theta $
is an intrinsic parameter of the soliton family which takes values $0<\theta
<\pi $, and 
\begin{eqnarray}
X &=&\kappa \left( 1-c^{2}\right) ^{-1/2}\left[ x-\xi (t)\right] ,\,T=\kappa
\left( 1-c^{2}\right) ^{-1/2}\left( t-cx\right) ,  \nonumber \\
\phi (X) &=&\frac{4c}{3-c^{2}}\mathrm{\tan }^{-1}\left\{ \tanh \left[ (\sin
\,\theta )X\right] \tan \left( \theta /2\right) \right\} ,  \label{params} \\
W(X) &=&\left( \sin \,\theta \right) \,\mathrm{sech}\left[ (\sin \,\theta
)X-i\left( \theta /2\right) \right] \,,  \nonumber
\end{eqnarray}
where $d\xi /dt=c$ \noindent (note the above-mentioned \textquotedblleft\
Lorentz contraction" of the soliton's size, obvious in the definition of $X$%
).

Equations (\ref{pdes}) conserve three dynamical invariants: the norm
(frequently called energy in optics) and momentum, 
\begin{equation}
E\equiv \int_{-\infty }^{-\infty }\left( \left\vert u\right\vert
^{2}+\left\vert v\right\vert ^{2}\right) dx,~P=i\int_{-\infty }^{+\infty
}\left( u_{x}^{\ast }u+v_{x}^{\ast }v\right) dx,  \label{EP}
\end{equation}
and also the Hamiltonian (which will not be used below). For the soliton
solution (\ref{movsol}), the norm and momentum take values 
\begin{eqnarray}
E_{\mathrm{sol}} &\equiv &\frac{8\theta \left( 1-c^{2}\right) }{3-c^{2}}, 
\nonumber \\
P_{\mathrm{sol}} &=&8\kappa c\sqrt{1-c^{2}}\left[ \frac{\left(
7-c^{2}\right) }{\left( 3-c^{2}\right) ^{2}}\left( \sin \theta -\theta \cos
\theta \right) +\frac{\theta \cos \theta }{3-c^{2}}\right]  \label{soliton}
\end{eqnarray}
(note that the energy does not depend on $\kappa $).

The expression for the momentum strongly simplifies in the \textquotedblleft
nonrelativistic" limit ($c^{2}\ll 1$), which makes it possible to identify
an effective soliton's mass, 
\begin{equation}
M_{\mathrm{sol}}\equiv \lim_{c\rightarrow 0}\frac{P_{\mathrm{sol}}}{c}=\frac{
8\kappa }{9}\left( 7\sin \theta -4\theta \cos \theta \right) .  \label{M}
\end{equation}
Note that the gap solitons are stable in the region $\theta \leq \theta _{ 
\mathrm{cr}}$, where $\theta _{\mathrm{cr}}$ is very close to $\pi /2$ [for
instance, $\theta _{\mathrm{cr}}\approx 1.01\left( \pi /2\right) $ if $c=0$] 
\cite{Tasgal}. In all the stability region, the mass (\ref{M}) increases
with $\theta $, i.e., larger $\theta $ corresponds to a \textquotedblleft
heavier" soliton.

\section{The adiabatic approximation}

In the apodized BG, with $\kappa =\kappa (x)$, the momentum is no longer
conserved; instead, the following exact equation can be derived from the
underlying equations (\ref{pdes}):

\begin{equation}
\frac{dP}{dt}=2\int_{-\infty }^{+\infty }\frac{d\kappa }{dx}\mathrm{Re}
\left( uv^{\ast }\right) dx.  \label{dP/dt}
\end{equation}
The adiabatic approximation applies to the case of a slowly changing $\kappa
(x)$, when it varies on a length which is essentially larger than the
soliton's size. As it follows from the expressions (\ref{params}) for the
soliton solution, this implies a general condition, $\sin \theta \gg \sqrt{
1-c^{2}}\left( \kappa ^{-1}\left\vert d\kappa /dx\right\vert \right) $.

For slowly varying $\kappa (x)$, we rewrite Eq. (\ref{dP/dt}) as 
\begin{equation}
\frac{dP}{dt}=2\frac{d\kappa }{dx}\int_{-\infty }^{+\infty }\mathrm{Re}
\left( uv^{\ast }\right) dx,  \label{slow_kappa}
\end{equation}
assuming that the value of $d\kappa /dx$ is taken at the point $x=\xi (t)$,
where the soliton's center is located at a given moment of time, see Eqs. (%
\ref{params}). Then, the left-hand side of Eq. (\ref{slow_kappa}) can be
calculated with the unperturbed soliton solution (\ref{movsol}), and the
expression for the soliton's momentum from Eqs. (\ref{soliton}) can be
substituted in the left-hand side. This yields the following general
evolution equation for the soliton's parameters [the velocity $c(t)$ and the
mass parameter $\theta (t)$]:

\begin{eqnarray}
&&\frac{d}{dt}\left\{ \kappa (\xi )c\sqrt{1-c^{2}}\left[ \frac{\left(
7-c^{2}\right) }{\left( 3-c^{2}\right) ^{2}}\left( \sin \theta -\theta \cos
\theta \right) +\frac{\theta \cos \theta }{3-c^{2}}\right] \right\} 
\nonumber \\
&=&-\frac{d\kappa }{d\xi }\frac{\left( 1-c^{2}\right) ^{3/2}}{3-c^{2}}\sin
\theta .  \label{general}
\end{eqnarray}
Note that, as $\kappa (x)$ is taken at the point $x=\xi (t)$, which changes
in time according to $d\xi /dt=c$, the coefficient $\kappa $ is subject to
the $t$-differentiation on the left-hand side of Eq. (\ref{general}).

Unlike the momentum, the energy of the wave fields remains the dynamical
invariant in the presence of the apodization. Therefore, using the
expression for the soliton's energy from Eqs. (\ref{soliton}), in the
adiabatic approximation one can eliminate $\theta (t)$ in favor of $c(t)$: 
\begin{equation}
\theta =\frac{3-c^{2}}{1-c^{2}}\frac{E}{8}.  \label{theta}
\end{equation}
Thus, Eqs. (\ref{general}) and (\ref{theta}), together with the
above-mentioned relations $\kappa =\kappa (\xi )$ and $d\xi /dt=c$,
constitute a closed system of the evolution equations for $\theta (t)$ and $%
c(t)$.

The adiabatic approximation strongly simplifies in the above-mentioned
``nonrelativistic" limit, $c^{2}\ll 1$: Eq. (\ref{theta}) then amounts to $%
\theta =\mathrm{const}$, the remaining equation for $c(t)$ being 
\begin{equation}
\frac{d}{dt}\left( \kappa c\right) =-\frac{3\sin \theta }{7\sin \theta
-4\theta \cos \theta }\frac{d\kappa }{d\xi }.  \label{nonrel}
\end{equation}
Further, using the identities $d\left( \kappa c\right) /dt\equiv \left[ d\left(
\kappa c\right) /d\xi \right] \left( d\xi /dt\right) =cd\left( \kappa
c\right) /dx\ $(as $d\xi /dt\equiv c$), we transform Eq. (\ref{nonrel}), 
\[
c\frac{d\left( \kappa c\right) }{d\xi }=-\frac{3\sin \theta }{7\sin \theta
-4\theta \cos \theta }\frac{d\kappa }{d\xi }, 
\]
which can be integrated to yield 
\begin{equation}
c^{2}(\xi )=\left( \frac{3\sin \theta }{7\sin \theta -4\theta \cos \theta}
+c_{0}^{2}\right) \frac{\kappa _{0}^{2}}{\kappa ^{2}(\xi )}-\frac{3\sin
\theta }{7\sin \theta -4\theta \cos \theta },  \label{res}
\end{equation}
where the subscript $0$ refers to initial values. In particular, Eq. (\ref%
{res}) implies that the soliton may be brought to a halt ($c^{2}=0$) at a
point $\xi _{\mathrm{halt}}$ where $\kappa ^{2}$ attains the value 
\begin{equation}
\kappa _{\mathrm{halt}}^{2}=\kappa _{0}^{2}\left[ 1+c_{0}^{2}\frac{\left(
7\sin \theta -4\theta \cos \theta \right) }{3\sin \theta }\right] .
\label{halt}
\end{equation}
Of course, the expression (\ref{halt}) makes sense if $\kappa _{\mathrm{halt}
}^{2}$ does not exceed the largest value of $\kappa ^{2}$ available in a
given apodization profile.

Note that, except for the special case when $\kappa ^{2}(\xi )=\kappa _{ 
\mathrm{halt}}^{2}$ is attained at $\xi =\infty $, Eq. (\ref{res}) predicts
that the soliton will not be stuck forever at the halt point $\xi _{\mathrm{%
\ halt}}$, but will actually bounce back. Indeed, with regard to the
relation $c=d\xi /dt$, it follows from Eq. (\ref{res}) that, around the halt
point, the soliton's law of motion takes the form $\xi _{\mathrm{halt}}-\xi
\sim \left( t-t_{\mathrm{halt}}\right) ^{2}$, where $t_{\mathrm{halt}}$ is
the moment of time at which the velocity $c(t)$ vanishes.

We also notice that, if the soliton passes an apodized region (step) across
which $\kappa (x)$ increases, then the final value of the velocity, as given
by Eq. (\ref{res}), increases with $\theta $. This complies with the above
conclusion that the soliton is heavier for larger $\theta $: due to the
inertia, a heavier object suffers smaller deceleration passing a potential
step.

\section{Numerical results}

The underlying equations (\ref{pdes}) were simulated with the step-wise
profile of the local reflectivity,

\begin{equation}
\kappa (x)=\kappa _{0}+\frac{1}{2}\Delta \kappa \tanh \left( \frac{x-x_{0}} {%
w}\right) ,  \label{stepk}
\end{equation}

\noindent where the step's height $\Delta \kappa $ is positive, and a
normalization condition is imposed, 
\begin{equation}
\kappa (=+\infty )=\kappa _{0}+\left( \Delta \kappa /2\right) \equiv 1.
\label{norm}
\end{equation}
The width of the step is $w$, and its center is located at the point $%
x=x_{0} $. The split-step fast-Fourier-transform method was used to run the
simulations.

In the simulations, the gap soliton was launched, as the exact solution (\ref%
{movsol}) with a positive velocity $c_{0}$, from the left edge of the
integration domain. Results will be presented, chiefly, for stable solitons,
which implies that the initial values of $\theta $ should be taken from the
interval $\theta _{0}\leq \pi /2$ (see above). In fact, simulations were
also performed for solitons with $\theta >\pi /2$. The results are not
drastically different for them (a brief description of this case is also
given below); in particular, the passage of the step does not catalyze the
onset of the soliton's instability, which develops on a larger time scale.

\subsection{Verification of the adiabatic approximation}

First of all, predictions produced by the above adiabatic approximation were
checked for the case of a sufficiently smooth step ($w$ not too small). The
checkup was done in two stages. In the lowest approximation, the explicit
solution (\ref{res}), that was derived in the \textquotedblleft
nonrelativistic" approximation, was used, along with the corresponding
condition $\theta =\mathrm{const}$. To obtain a more accurate result at the
second stage, the full evolution equation (\ref{general}), combined with the
relation (\ref{theta}), was solved numerically. It was concluded that the
adiabatic approximation is quite accurate, as long as the step remains
smooth. This can be illustrated by typical results obtained for the soliton
with the initial value of its intrinsic parameter $\theta _{0}=\pi /2$
impinging on the step with the width $w=6$.

In this case, the special value of the initial velocity $\left( c_{0}\right)
_{\mathrm{halt}}$, which provides for the halt of the soliton in the direct
simulations, was sought for at fixed values of $\kappa _{0}$ (the ``halt"
was realized so that the soliton's velocity would drop to zero, and after
being stuck for a long time, the soliton would then very slowly start to
move backward). It was thus found that the numerical values are, for
instance, 
\begin{equation}
\left( c_{0}\right) _{\mathrm{halt}}\left( \kappa _{0}=0.8\right)
=0.45;~\left( c_{0}\right) _{\mathrm{halt}}\left( \kappa _{0}=0.5\right)
=0.74.  \label{c_halt}
\end{equation}
On the other hand, for these values of $c_{0}$ the adiabatic approximation
predicts which must be the final values of the reflectivity, $\kappa _{ 
\mathrm{final}}\equiv \kappa _{0}+(1/2)\Delta \kappa $ [see Eq. (\ref{stepk}
)], that will halt the soliton. In particular, in the \textquotedblleft
nonrelativistic" approximation, we simply set $c^{2}=0$ in Eq. (\ref{res}),
which yields 
\begin{equation}
\kappa _{\mathrm{final}}^{2}=\kappa _{0}^{2}\left( 1+\frac{7\sin \theta
-4\theta \cos \theta }{3\sin \theta }c_{0}^{2}\right) .  \label{nonrel_comp}
\end{equation}
The predicted value of $\kappa _{\mathrm{final}}$ must be compared with 
$\kappa _{\mathrm{final}}\equiv 1$, which was imposed by the normalization (%
\ref{norm}) adopted in the full numerical simulations.

The ``nonrelativistic" approximation predicts, for the two cases indicated
in Eq. (\ref{c_halt}), $\kappa _{\mathrm{final}}=0.971$ and $0.755$,
respectively. The former value is quite close to the exact one, $\kappa _{%
\mathrm{final}}\equiv 1$. The latter value is not very close to it, but the
corresponding initial velocity, $c_{0}=0.74$, does not correspond to the
``nonrelativistic" case. On the other hand, the use of the full
adiabatically derived equation (\ref{general}) predicts, for the same two
cases, $\kappa _{\mathrm{final}}=1.017$ and $1.023$, respectively.

\begin{figure}[tbh]
\centering \includegraphics[width=4in]{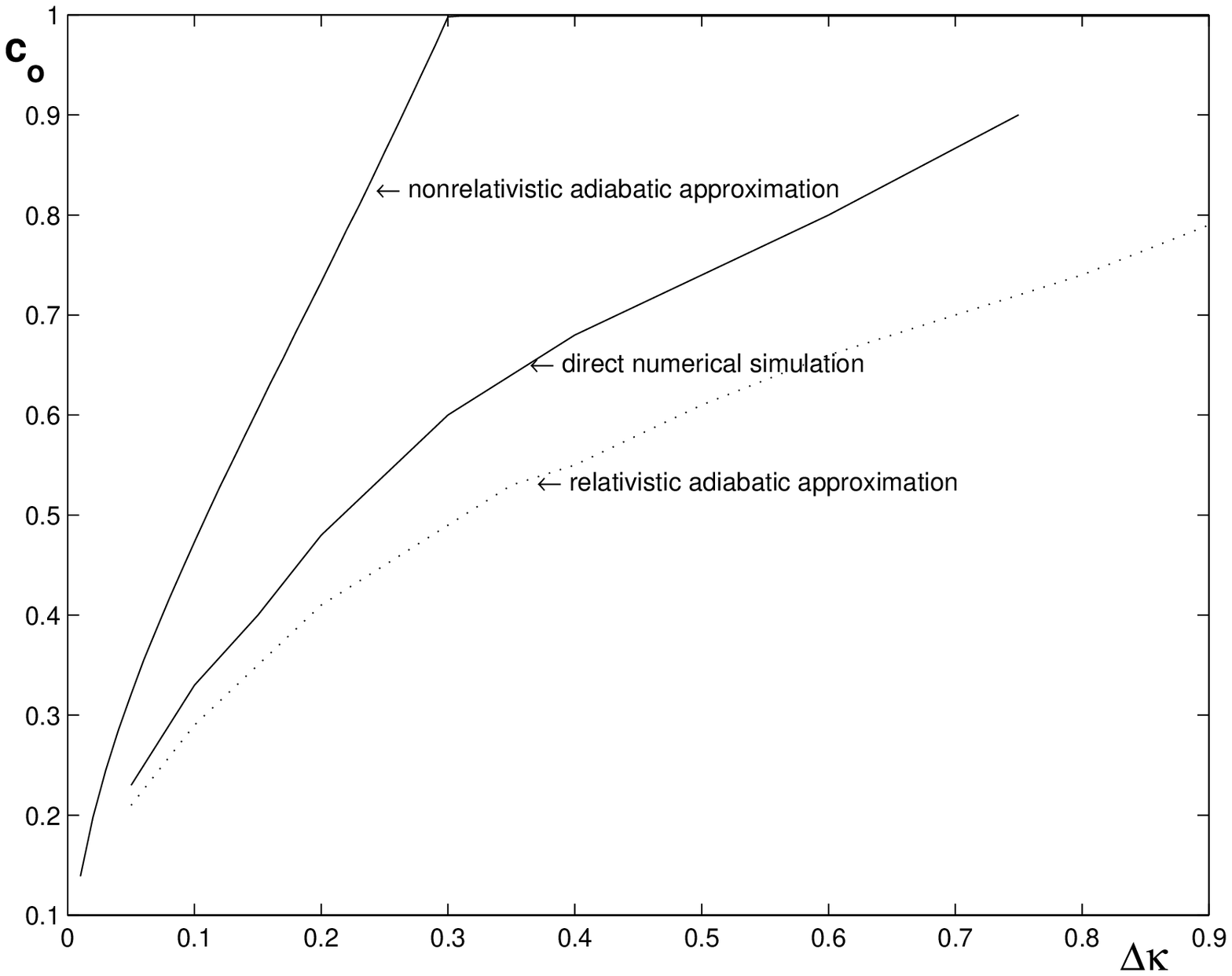}
\caption{Borders between the bounce of the gap soliton, initially moving at
the velocity $c_{0}$, from the local-reflectivity step with the height of $%
\Delta \protect\kappa $ [see Eq. (\protect\ref{stepk})], and its passage
over the step, are shown in the parameter plane $(\Delta \protect\kappa %
,c_{0})$, as found from direct simulations, and from the analytical
approximations -- the full (``relativistic") one, based on Eqs. (\protect\ref%
{general}) and (\protect\ref{theta}), and its simplified
(``nonrelativistic") version, which is based on Eq. (\protect\ref{res}) and
assumes $c_{0}^{2}\ll 1$ and $\protect\theta =\mathrm{\ const}$. The results
are presented for the fixed step's width, $w=20$, and fixed mass parameter
of the impinging soliton, $\protect\theta =\protect\pi /2$. Both $\Delta 
\protect\kappa$ and $c_{0}$ are in normalized units. }
\label{Fig1}
\end{figure}

Systematic comparison of the numerical and analytical results is presented
in Fig. \ref{Fig1}, which shows the border between the reflection of the
moving gap soliton from the apodization step, and its passage through the
step, in the plane of ($c_{0}$,$\Delta \kappa $). For smaller values of the
step's size, the analytical approximation provides for results which are
quite close to the numerical findings. Making the perturbation stronger
(increasing $\Delta \kappa $), we observe a growing deviation, but the full
``relativistic''\ approximation, based on Eqs. (\ref{general}) and (\ref%
{theta}), still provides for a reasonable approximation even for larger $%
\Delta \kappa $ (and larger velocity), while the oversimplified
``nonrelativistic''\ approximation, based on the single equation (\ref{res}%
), becomes irrelevant in that case.

\subsection{Different outcomes of the collision of the gap soliton with the
apodization step}

To present direct numerical results in a systematic form (especially, for
the situation with a steeper apodization profile), we first fix the height
and width of \ the step, setting $\Delta \kappa =0.5$ and $w=6$ [note that
this width is much smaller than it was in the case of Fig. \ref{Fig1} ($w=20$%
), hence the step is much steeper in the present case]. Launching the
solitons with various initial values $c_{0}$ and $\theta _{0}$ of the
velocity and mass parameter to hit the apodization step, it was found that,
as it might be naturally expected, more energetic solitons, with a large
velocity $c$ and/or larger mass parameter $\theta $, pass the step with
deceleration, a typical example of which is shown in Fig. 2(a),\ and slower
(smaller $c$) and/or lighter (smaller $\theta $) solitons bounce back from
the step, see a typical example in Fig. 2(b). Note that the only difference
between the cases displayed in Figs. 2(a) and 2(b) is that $c_{0}=0.75$ in
the former case, and $c_{0}=0.7$ in the latter case, showing that the border
between the transmission and repulsion is in between these values. On the
other hand, the calculation based on the quasi-particle equation of motion (%
\ref{general}) predicts that the border is at $c=c_{0}\approx 0.61$ for the
same case, which agrees reasonably well with the numerical results,
considering that the apodization profile is rather steep, and the
apodization step is high. The applicability of the adiabatic approximation
to the present cases is also corroborated by the observation that very
little radiation loss was generated in the direct simulations.

If the bounce takes place at a point $\xi _{\mathrm{halt}}$ located far to
the right from the step, the soliton gets stuck there for a finite but long
time, also in agreement with the prediction of the quasi-particle
approximation. An example of such a case (which, obviously, is of special
interest to the experiment and applications, suggesting a real possibility
to capture a pulse of ``standing light") is shown in Fig. 2(c). Note that,
in comparison with a more generic case of the ricochet shown in Fig. 2(b),
in the case of the quasi-trapping of the gap soliton, the bounce indeed
occurs after the soliton has advanced much farther to the right.

In the same situation as considered above, but with smaller values of $%
\theta _{0}$, the quasi-particle approximation does not apply, as the
soliton size becomes much larger than the step's width, hence the step may
not be considered as a smooth one in any approximation. The simulations
indeed produce a drastically different result in this case: hitting the
step, the soliton \emph{splits} into two pulses, transmitted and reflected
ones, which is illustrated by Fig. 2(d). In this case, both pulses
eventually decay into radiation, rather than self-trapping into
small-amplitude solitons.

The controllable splitting of the incident soliton is an interesting effect
in its own right, and it has a potential for various applications. Splitting
of the BG soliton hitting a chirped fiber grating has already been observed
experimentally and reproduced in numerical simulations \cite{express}. In
fact, in that case the incident soliton split into three pulses: a
transmitted soliton and two reflected ones.

In this connection, it is relevant to mention that, although we do not
display detailed results for the (generally) unstable solitons with $\theta
>\pi /2$, the interaction with the step was simulated for them too. In most
cases, the intrinsic instability of the soliton does not set it in the
course of the limited time before it hits the step (which corresponds to the
experimental situation, in which the segment of the fiber grating before the
apodized region is not going to be very long \cite{experiment}). Then, if
the step is smooth, the heavy soliton passes it or bounces back, without
catalyzing the onset of the intrinsic instability, and without conspicuous
emission of radiation. The soliton emits an appreciable amount of radiation
if the barrier is steeper. The splitting of the soliton with $\theta >\pi /2$
into forward and backward moving parts, similar to what is shown for smaller 
$\theta $ in Fig.~2(d), does not occur on the barrier with $w=6$, which is
the value selected for Fig. 2, suggesting that the soliton is a more
cohesive object for larger $\theta $ (see further discussion of this point
below). Splitting takes place if the barrier is made still steeper and
taller -- for instance, with $w=4$ and $\Delta \kappa =0.7$, see Fig. 3. As
well as in the case displayed in Fig. 2(d), in the latter case the two
pulses do not eventually self-trap into secondary solitons, but rather decay
into radiation.

\subsection{Scanning the parameter space}

Having displayed typical examples of different outcomes of the collision, we
proceed to summarize the results in a systematic form. First, a
comprehensive description of the transition from the bounce of the soliton
to the passage, together with the splitting (if any), are provided, for
different values of the barrier's height $\Delta \kappa $, by the plots in
Fig. 4. For fixed $\theta $, they show the share $\varepsilon $ of the
initial soliton's energy which is reflected back as a result of the
interaction of the soliton with the step. A steep drop from $\varepsilon =1$
to $\varepsilon =0$ in Figs. 4(b) and 4(c) implies the transition from the
bounce to the transmission without splitting. On the contrary, a gentle
crossover means that the impinging soliton was split into two parts, with
the energy ratio between them depending on the initial velocity. These plots
clearly show strengthening of the soliton's integrity (its stabilization
against the splitting) with the increase of its energy.

Further, in Fig. 5 we show the velocities $c_{f}$ and $c_{b}$ of the
transmitted (``final") and reflected (bouncing) fragments of the incident
soliton, together with the backscattered-energy share $\varepsilon $, as
functions of $\Delta \kappa $. In the ranges where only one velocity is
shown in Fig. 5, the soliton is transmitted or reflected as a whole, while
the splitting takes place in intervals of the values of $\Delta \kappa $
where both continuous curves (for $c_{f}$ and $c_{b}$) coexist in Fig. 5.

Note that the transition from the full transmission to full reflection,
i.e., from $\varepsilon =0$ to $\varepsilon =1$, takes place in exactly the
same interval where the splitting occurs, cf. Fig. 4. As is seen, the
splitting interval is very narrow for the heavier soliton, with $\theta
=0.5\pi $, and much broader for the lighter one, with $\theta =0.3\pi $,
which is another common feature with the dependences displayed in Fig. 4.

The dependence of the outcome of the collision on the width of the
apodization region was studied too, as it helps to understand how really
wide the barrier must be to secure the adiabatic character of the
interaction, and how narrow it is when the splitting occurs. In Fig. 6, we
show the dependence of the final velocity of the transmitted soliton, $c_{f}$%
, and the reflected-energy share, $\varepsilon $, on the width while the
step's height is fixed, $\Delta \kappa =0.5$. The final velocity of the
transmitted soliton approaches an asymptotic value for large values of $w$.

Figure 6 shows that the transition from the strongly non-adiabatic regime
(with splitting) to a nearly adiabatic one is steep itself, taking place in
a relatively narrow interval around $w=3$. A practically important
implication of the results presented in Fig. 6 is that, for the
experimentally relevant case, with the intrinsic size of the BG soliton $%
\sim 1$ mm \cite{experiment}, the step which may be regarded as a
sufficiently smooth one should have the width $_{\sim }^{>}~3$ mm.

\section{Dynamics of the gap soliton in a periodically-apodized structure}

To test the robustness of the gap solitons in the fiber gratings with a
variable Bragg reflectivity $\kappa $, and its potential for the use in more
sophisticated devices, we also simulated the motion of the soliton in the
grating subjected to a periodic modulation of $\kappa $. It was built as a
periodic concatenation of steps with alternating signs of $\Delta \kappa $;
the fiber grating of this type may be considered as a Bragg
``superstructure". This scheme can be realized either directly in a long
periodically modulated grating, or as an apodized BG written in a \emph{%
fiber loop}, although the stability problems in these two settings are not
equivalent. The soliton dynamics in such a superstructure is a problem of
interest in its own right -- cf. the study of solitons in other periodic
strongly inhomogeneous nonlinear optical media, such as fibers with
dispersion management \cite{DM}, various schemes with ``nonlinearity
management" \cite{nonlin_management} (including layered media with a
periodically changing sign of the Kerr nonlinearity \cite{layered}),
``tandem" structures \cite{tandem}, the ``split-step" model \cite{SSM},
waveguide-antiwaveguide concatenations \cite{Gisin}, etc. A unifying feature
of these systems is the surprising robustness of the solitons in them.

Typical results produced by the simulations of the soliton propagation in
the periodic ``superstructure" are displayed in Fig. 7. It can be seen that
the soliton travels slower in segments with larger $\kappa$, and faster in
those with smaller $\kappa $. In the case when the width $w$ of each step is
relatively small [for instance, $w=1$, see panel 7(a)], the soliton
evolution is clearly non-adiabatic, cf. the above results for the single
step (we stress that the full modulation period in the present case is
essentially large than $w$, see Fig. 7). Accordingly, the soliton gradually
develops distortions in its shape and simultaneously emits radiation waves.
Parallel to distorting itself, the soliton also develops random fluctuations
of its velocity.

In the case when the periodic structure is composed of steps with a larger
width, for instance, $w=10$ [see Fig. 7(b)], the soliton also accumulates
distortion and generates some radiation loss, but, in accordance with the
expectation that the evolution must be close to adiabatic in this case, this
happens much slower, and the distortion remains mild after passing a long
distance. With a still larger step width (for instance, $w=15$), no
accumulation of distortion or radiation could be seen at all (the latter
case is not shown here, as one would simply observe an unperturbed soliton).
So, large values of $w$ indeed provide for the adiabatic character of the
soliton's motion through the periodic structure, but the smallest size of $w$
which makes it possible in the case of the periodic system is (quite
naturally) much larger than the minimum width which provided for the
adiabatic passage of the gap soliton through the single step, cf. Fig. 6.

Lastly, it is relevant to mention that virtually identical results were
observed in the simulations of the periodic system either with periodic
boundary conditions (b.c.) in $x$ (which corresponds to the above-mentioned
case of an apodized BG written in a fiber loop) or just in a long domain
with absorbing b.c. Thus, the same soliton dynamics is expected in both
above-mentioned physical realizations of the periodic \textquotedblleft
Bragg superstructure".

\section{Conclusion}

In this work, we have studied in detail motion of gap solitons in two models
of apodized fiber Bragg gratings, including, respectively, a single step of
the local reflectivity, or a periodic ``Bragg superstructure" consisting of
alternating steps with opposite signs. Both structures offer a potential for
the design of various optical elements employing the gap (Bragg) solitons.
The most important implication of the considered problem is a possibility to
halt the soliton by passing it through a step with an increasing
reflectivity, and eventually to capture a pulse of \emph{standing light}.

We have developed an analytical approach, which assumes adiabatic evolution
of the soliton, and is based on the balance equation for its momentum and
conservation of the energy. The result of the analysis can be obtained in a
fully explicit form in the ``nonrelativistic" case, and in an implicit form
(as an ordinary differential equation) in the general case. Comparison of
the predictions produced by the approximation with direct simulations shows
good accuracy, provided that the width of the inhomogeneity is essentially
larger than the soliton's intrinsic width.

Results of systematic direct simulations of the soliton's motion were
summarized, showing that the soliton is either transmitted across the step
or bounces from it, provided that the step is not too narrow; in particular,
it is possible to halt the soliton for a very long time. If the step is
narrow (so that the interaction of the incident soliton with it is no longer
adiabatic), the soliton may split into two pulses, transmitted and reflected
ones (the splitting of a gap soliton in a chirped fiber grating has already
been observed in the experiment \cite{express}). We have studied in detail
dependences of the outcome of the interaction on the height and width of the
step, as well as on the initial parameters (velocity and effective mass) of
the soliton. In particular, a general conclusion is that the soliton is a
more cohesive object, being less prone to the splitting, if its energy
(effective mass) is larger.

Moving across the periodic structure, the soliton accumulates distortion and
radiation loss if the structure is composed of narrow steps. The
perturbations accumulate much slower if the steps are wider, and in the
system composed of sufficiently wide steps the soliton can move without any
loss or irreversible deformation.

\section*{Acknowledgement}

We appreciate a valuable discussion with C.M. de Sterke. One of the authors
(B.A.M.) appreciates hospitality of the Optoelectronic Research Centre at
the Department of Electronic Engineering, City University of Hong Kong.

\newpage

\section*{Figures}

\begin{figure}[tbh]
\centering \includegraphics[width=4in]{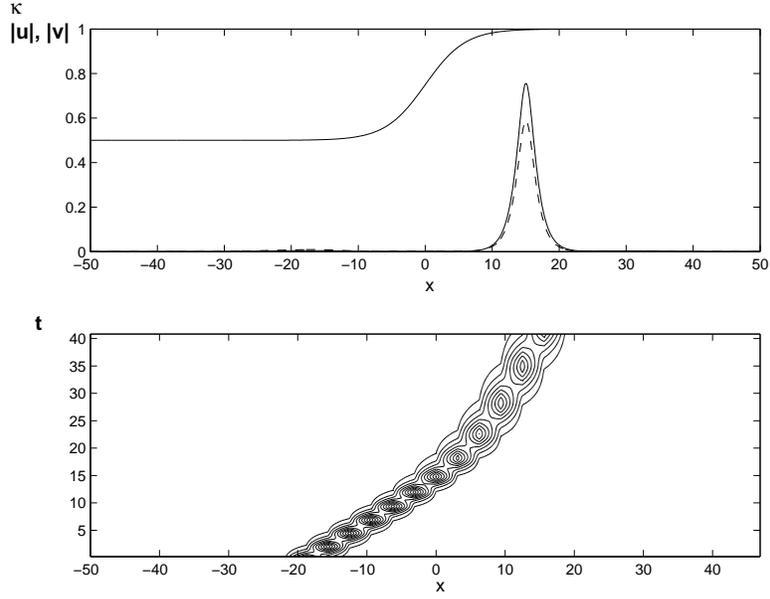}
\caption{Different typical outcomes of the collision of the gap soliton with
the apodization step that has the hight $\Delta \protect\kappa =0.5$ and
width $w=6$. The upper panels show the waveforms $|u(x)|$ and $|v(x)|$
(solid and dashed lines) at the end of the simulation, together with the
apodization profile, $\protect\kappa (x)$. The lower panels depict the
evolution of the field $|u(x)|$ in terms of level contours. (All the
variables plotted are in normalized units.) (a) Deceleration of the soliton
with very little radiation loss. The initial parameters of the soliton are $%
c_{0}=0.75$ and $\protect\theta _{0}=0.5\protect\pi $.}
\label{Fig2a}
\end{figure}

\addtocounter{figure}{-1}

\begin{figure}[tbh]
\centering \includegraphics[width=4in]{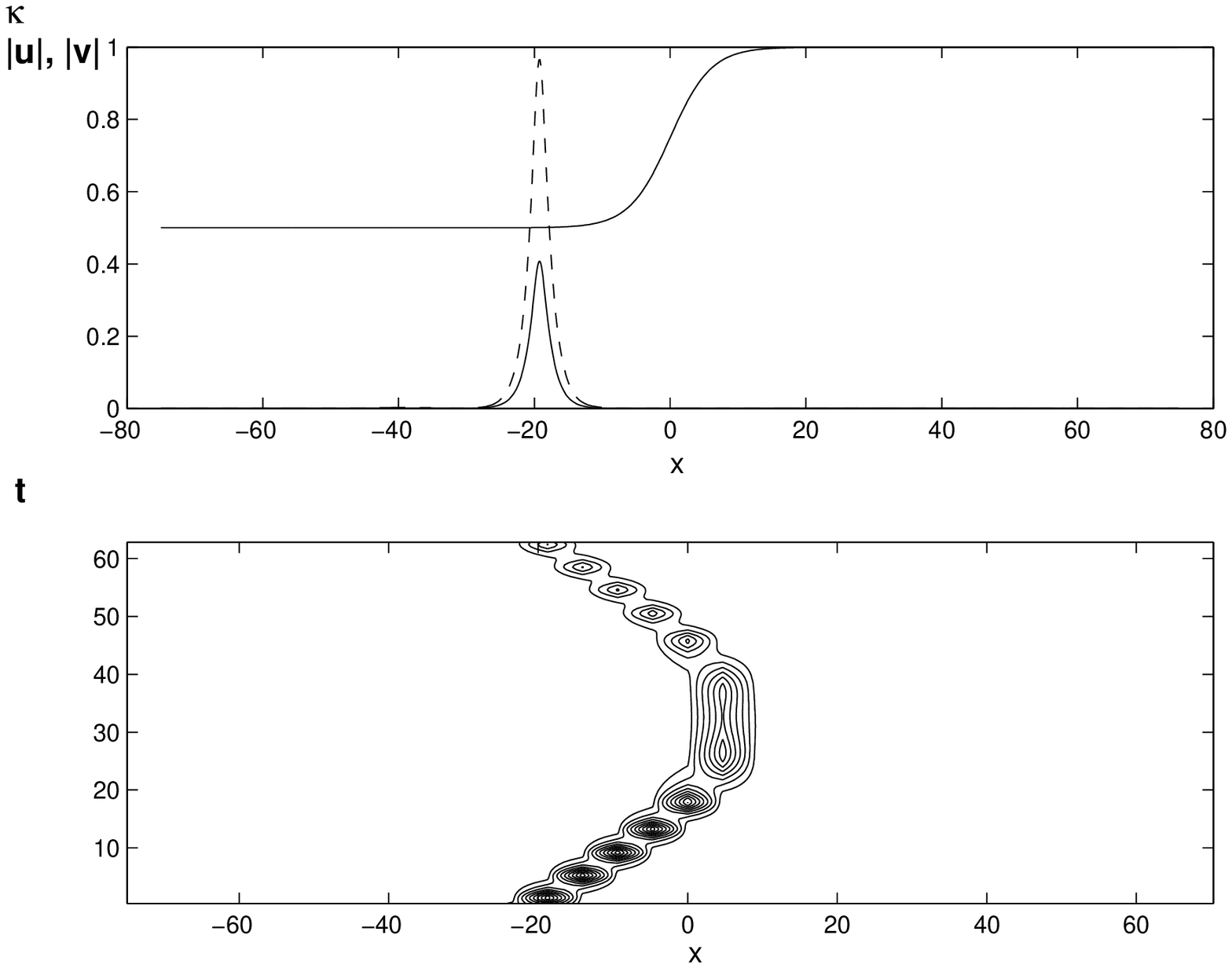}
\caption{ (b) The soliton bouncing back with almost no radiation loss. Here, 
$c_{0}=0.70$ and $\protect\theta _{0}=0.5\protect\pi $.}
\label{Fig2b}
\end{figure}

\addtocounter{figure}{-1}

\begin{figure}[tbh]
\centering \includegraphics[width=4in]{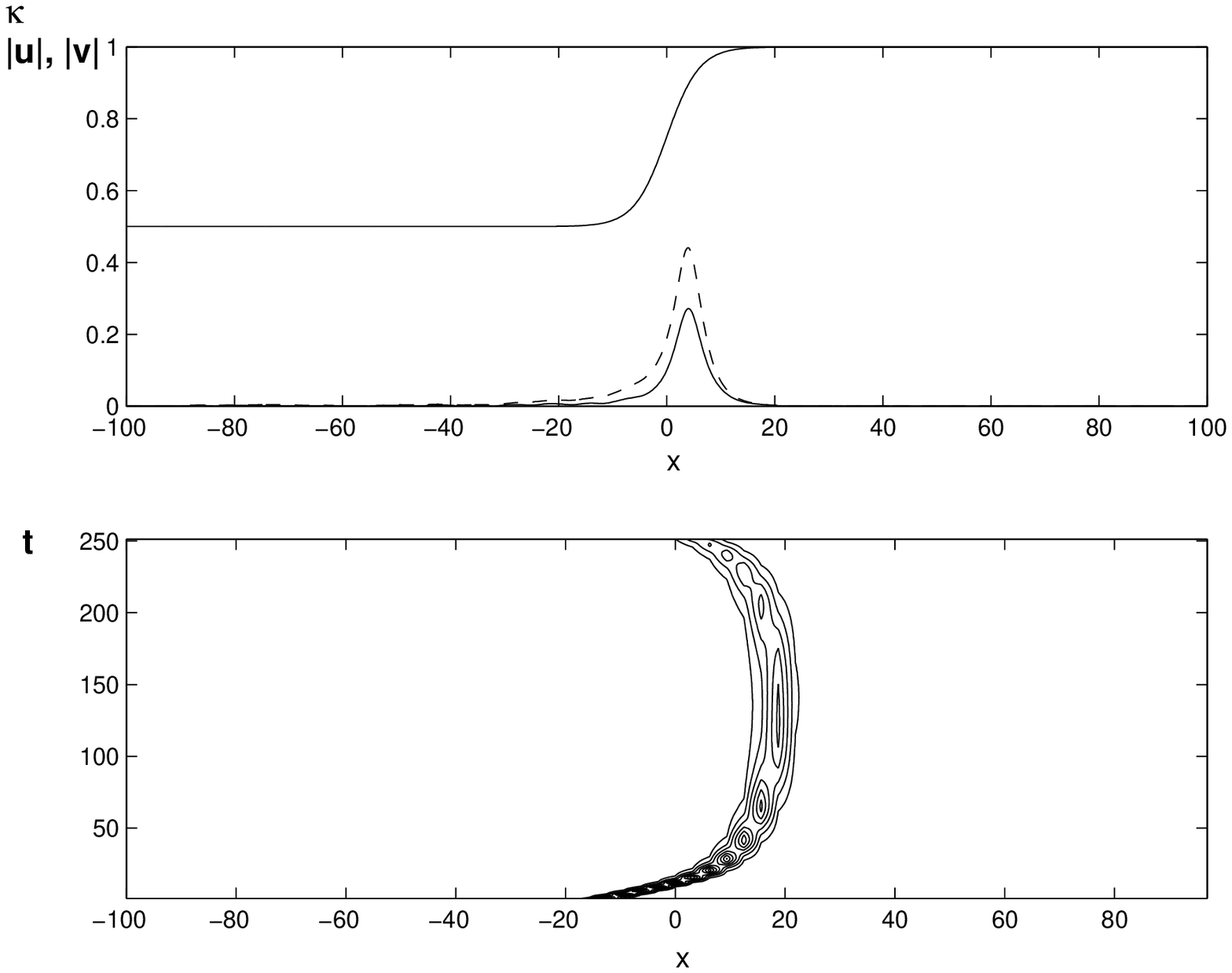} 
\caption{ (c) An example of the soliton being halted by the step and staying
immobile for a long time. In this case, $c_{0}=0.7403$ and $\protect\theta %
_{0}=0.5\protect\pi $.}
\label{Fig2c}
\end{figure}

\addtocounter{figure}{-1}

\begin{figure}[tbh]
\centering \includegraphics[width=4in]{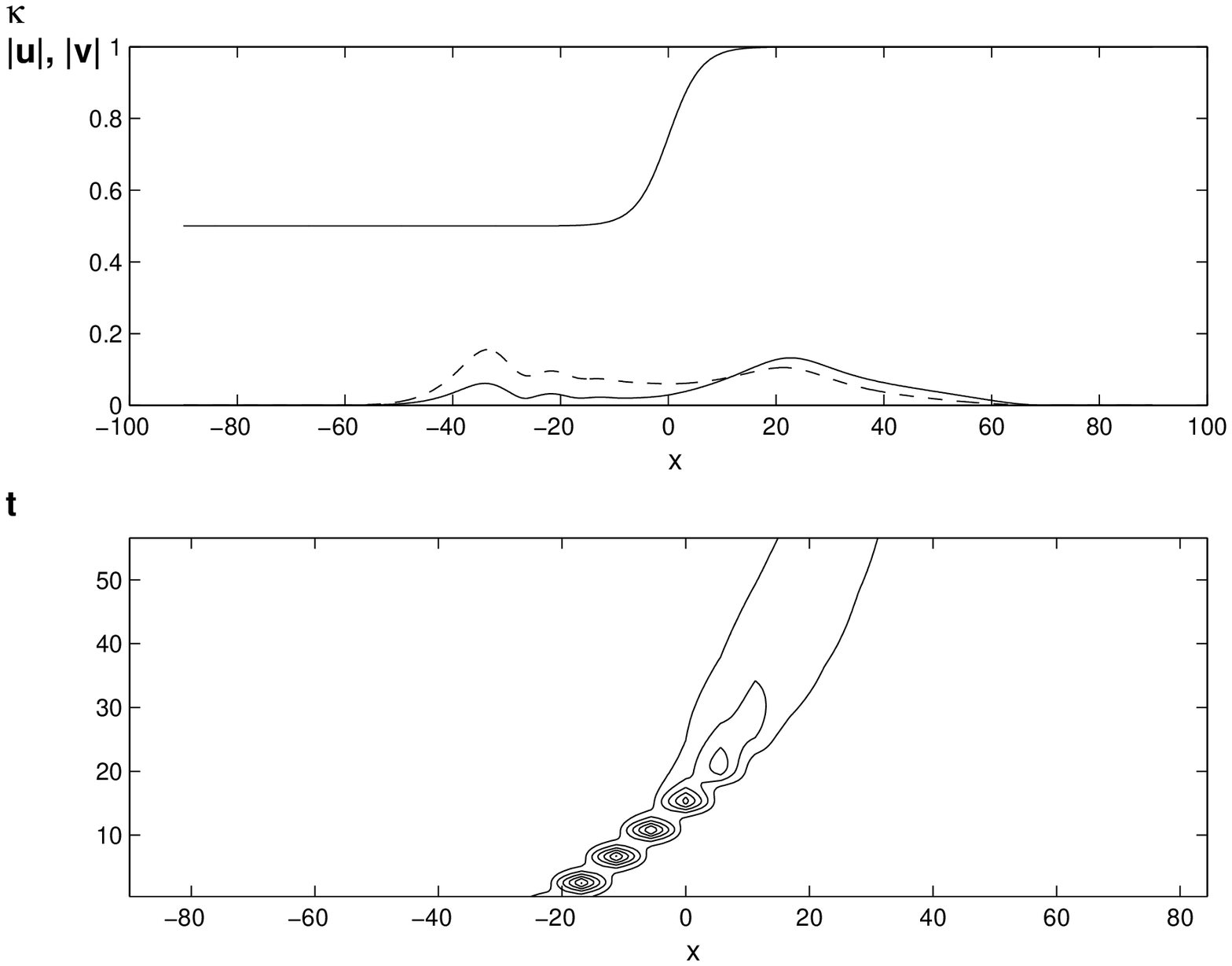}
\caption{ (d) Split of the soliton into two pulses, transmitted and
reflected ones. In the course of subsequent evolution, both pulses gradually
decay into radiation. In this case, $c_{0}=0.8$ and $\protect\theta _{0}=0.3 
\protect\pi $.}
\label{Fig2d}
\end{figure}
\bigskip

\begin{figure}[tbh]
\centering\includegraphics[width=4in]{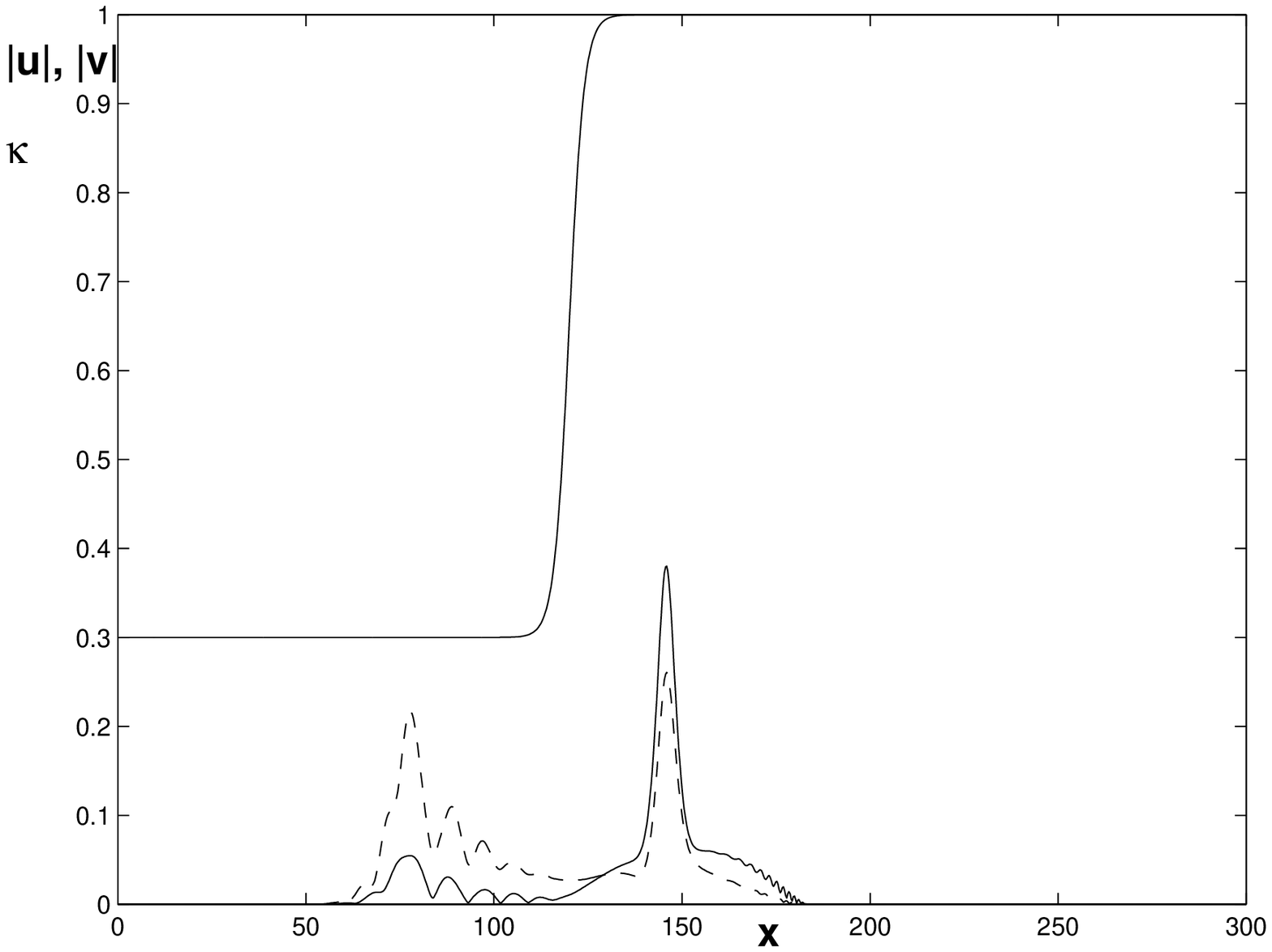}
\caption{The same as in Fig. 2(d), in the case of a narrower and taller
apodization step, with $w=4$ and $\Delta \protect\kappa =0.7$, for the
impinging soliton whose mass parameter exceeds the instability threshold, $%
\protect\theta =0.6\protect\pi $. All the variables plotted are in
normalized units.}
\label{Fig3}
\end{figure}

\begin{figure}[tbh]
\centering\includegraphics[width=4in]{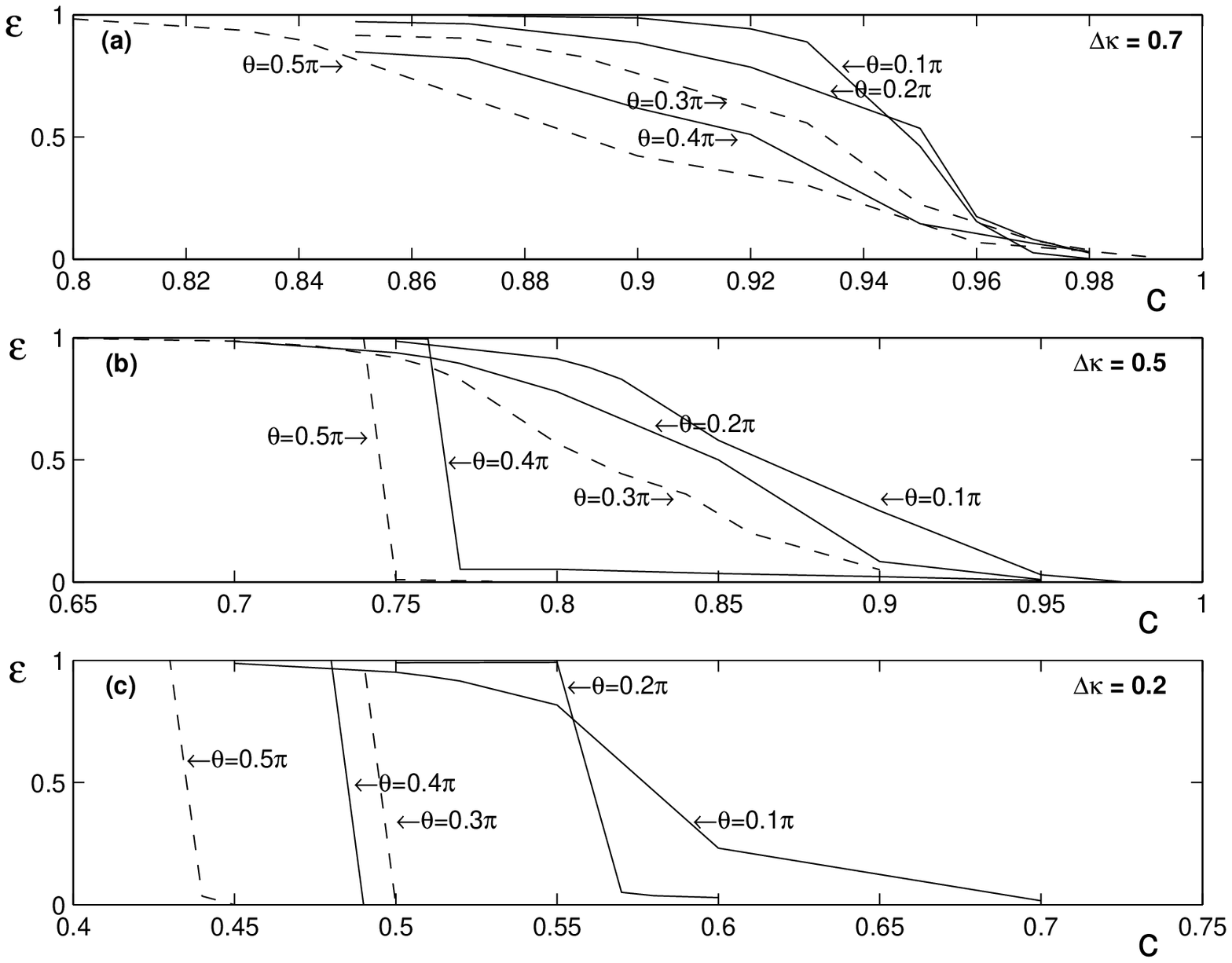}
\caption{The share of the initial soliton's energy, which is reflected back
by the apodization step of the width $w=6$, versus the initial velocity $c$,
for different values of the initial mass parameter $\protect\theta $, and
different values of the step's height $\Delta \protect\kappa $. Three panels
of the figure show the dependence $\protect\varepsilon (c)$ in different
regions of $c$ and on different scales, in order to highlight regions where
nontrivial changes occur. The steep drop of $\protect\varepsilon $ from $1$
to $0$ for larger $\protect\theta $ (such as $\protect\theta =0.4$ and $0.5$%
) implies the transition from the bounce of the soliton to the transmission
without splitting. ($\protect\varepsilon$ is dimensionless, and $c$ is in
normalized unit.)}
\label{Fig4}
\end{figure}
\bigskip

\begin{figure}[tbh]
\centering \includegraphics[width=4in]{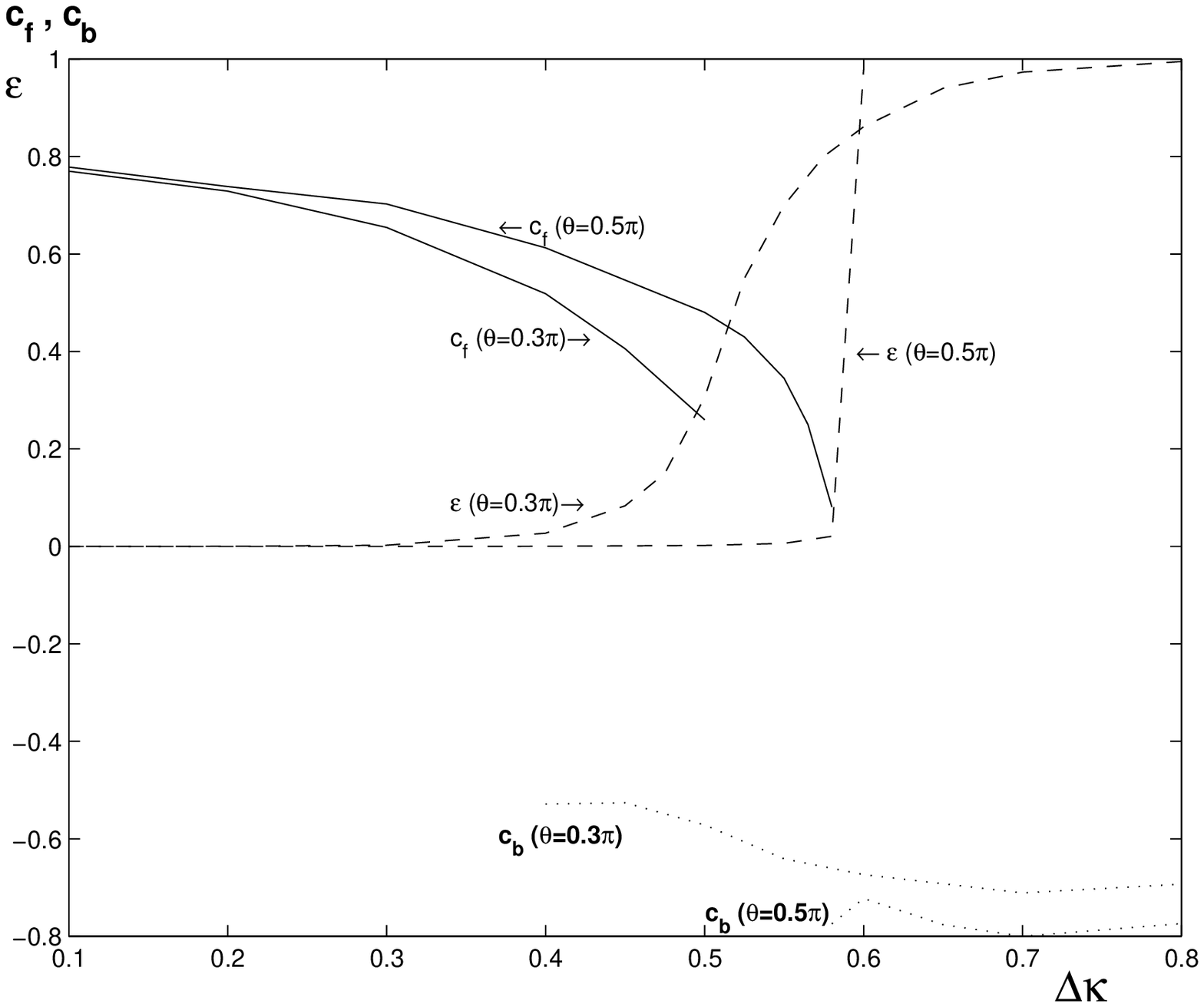}
\caption{The final velocity $c_{f}$ of the transmitted pulse and the
velocity $c_{b}$ of the reflected one (solid lines) are shown, together with
the backscattered-energy share $\protect\varepsilon $, versus the height of
the apodization step $\Delta \protect\kappa $, for $\protect\theta =0.3 
\protect\pi $ and $\protect\theta =0.5\protect\pi $. The splitting of the
incident soliton takes place in the interval where both solid curves are
present. In this figure, the step's width and initial velocity are fixed, $%
w=6$ and $c_{0}=0.8$. ($\protect\varepsilon$ is dimensionless, and $c_{f}$, $%
c_{b}$, $\Delta \protect\kappa $ are in normalized units.)}
\label{Fig5}
\end{figure}
\bigskip

\begin{figure}[htb]
\centering \includegraphics[width=4in]{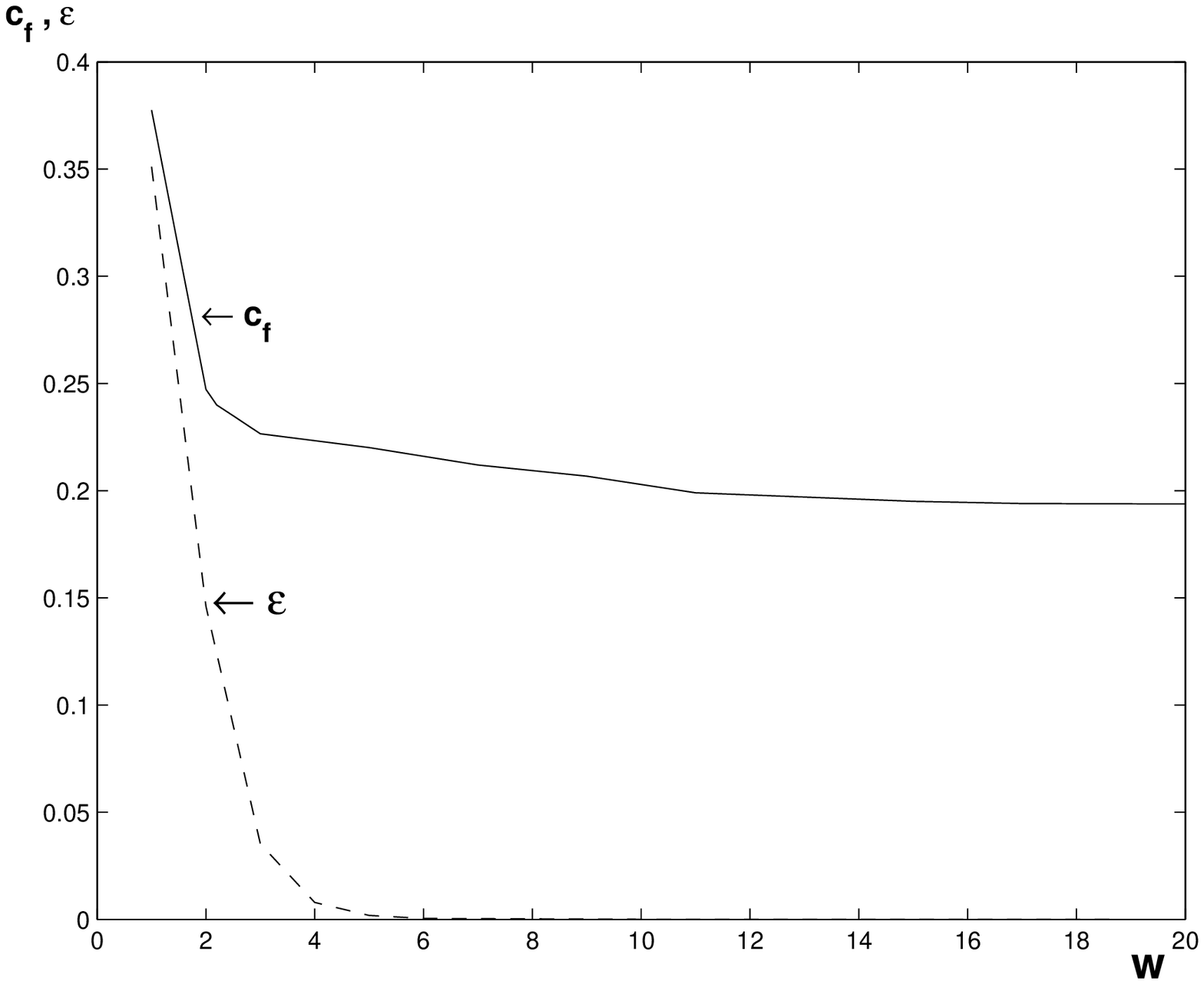}
\caption{The final soliton's velocity $c_{f}$ (solid line) and the fraction
of the backscattered energy, $\protect\varepsilon$ (dashed line), versus the
apodized-layer's width $w$, in the case when the height of the step is $%
\Delta\protect\kappa =0.5$. In this figure, the initial velocity and mass
parameter of the soliton are fixed to be $c_0=0.75$ and $\protect\theta_0=
0.5\protect\pi$. ($\protect\varepsilon$ is dimensionless, and $c_{f}$, $w$
are in normalized units.)}
\label{Fig6}
\end{figure}
\bigskip

\begin{figure}[tbh]
\centering \includegraphics[width=4in]{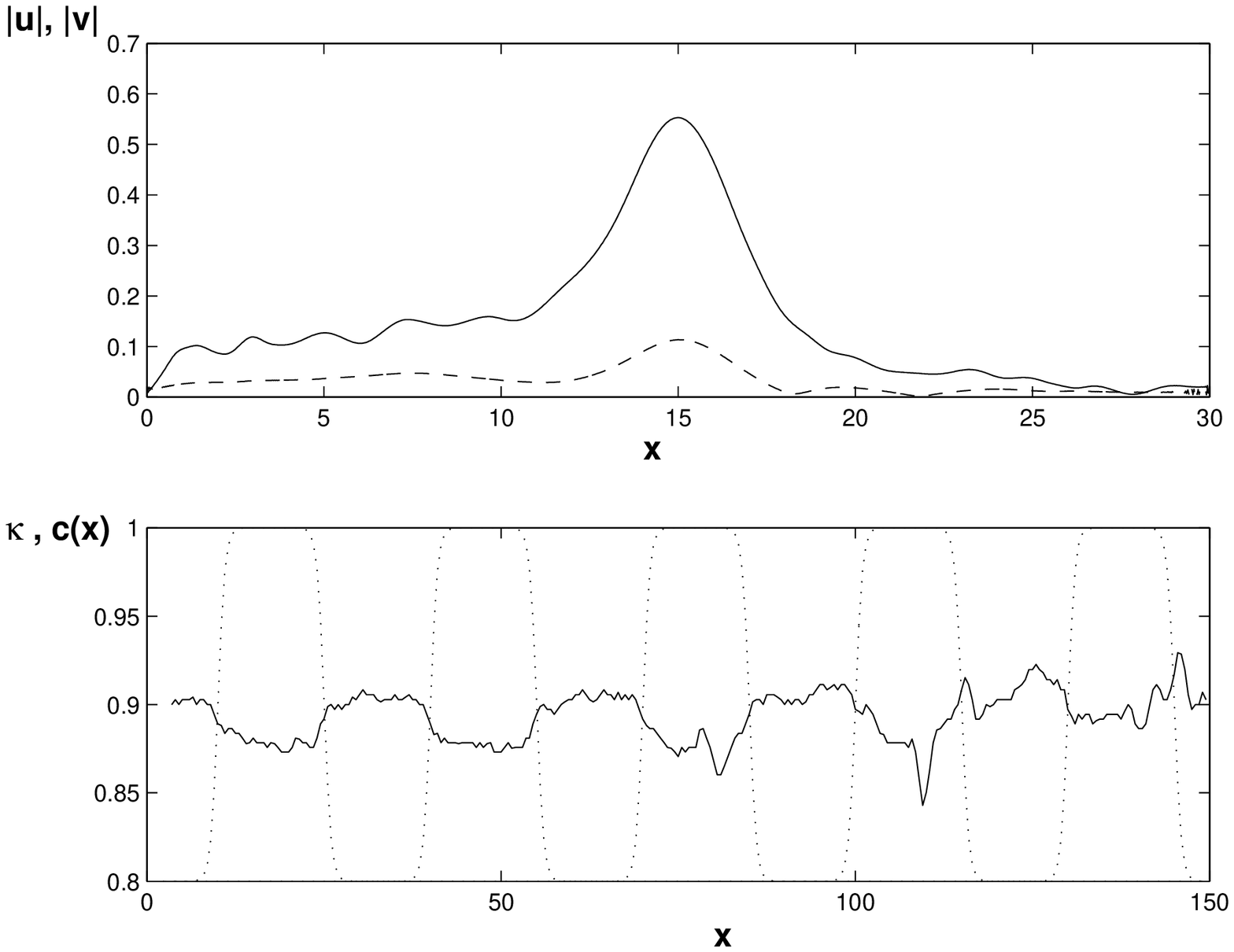}
\caption{Motion of the soliton with the initial velocity $c_{0}=0.9$ and
initial mass parameter $\protect\theta _{0}=0.5\protect\pi $ through a
periodic structure consisting of alternating steps with the local Bragg
reflectivity $\protect\kappa $ varying between $0.8$ and $1$. Bottom
portions in each panel show the profile of $\protect\kappa (x)$ (dotted
line) and evolution of the soliton's velocity (continuous line). The top
portions show the shape of the soliton ($|u(x)|$ and $|v(x)|$) at the end
of the long simulation. (All the variables plotted are in normalized units.)
(a): The width of the individual step is $w=1$ (a non-adiabatic case).}
\label{Fig7a}
\end{figure}
\bigskip

\addtocounter{figure}{-1}

\begin{figure}[htb]
\centering \includegraphics[width=4in]{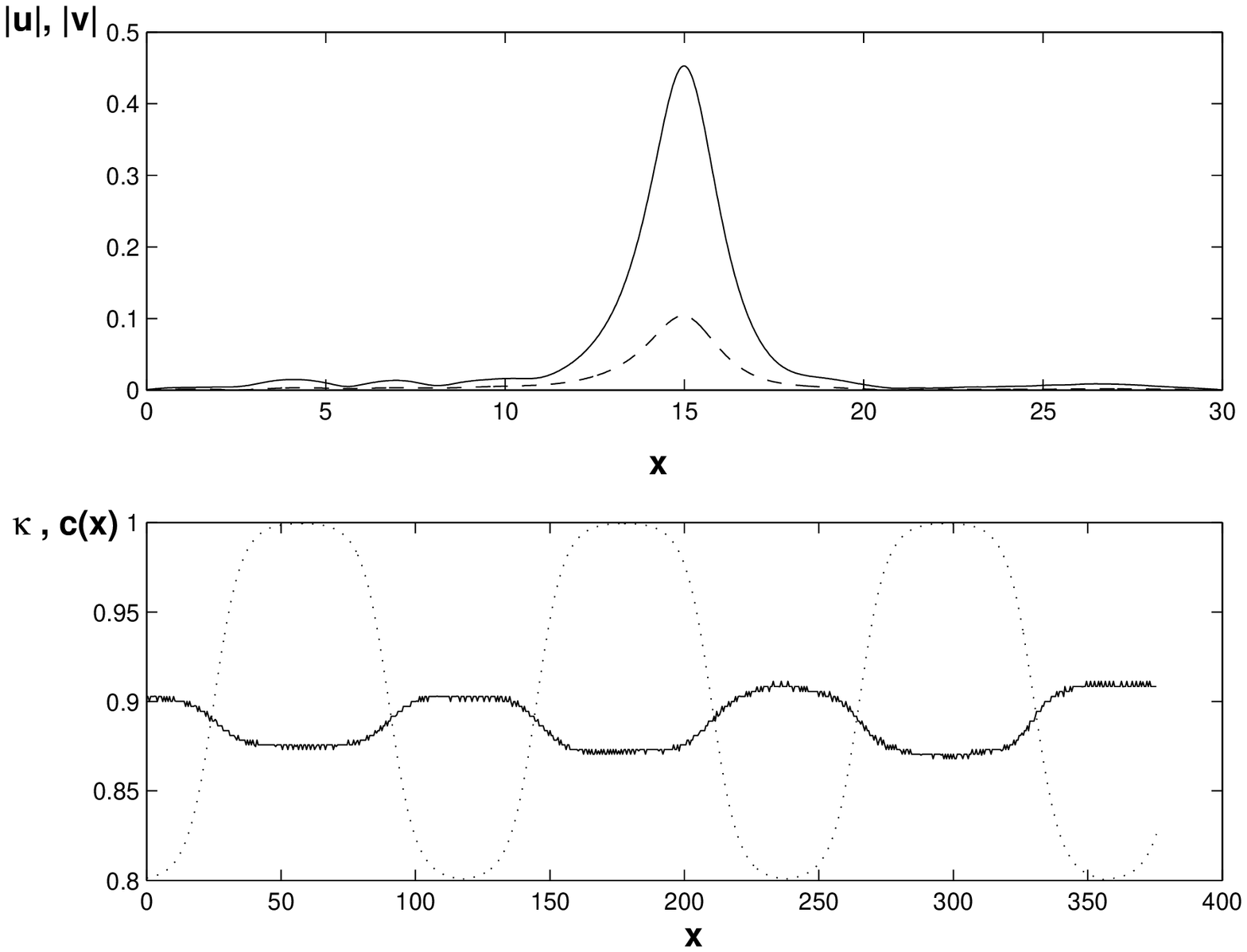}
\caption{(b): The width of each step is $w=10$ (a nearly adiabatic case).}
\label{Fig7b}
\end{figure}
\bigskip

\end{document}